\documentclass[aps,prl,preprint,groupedaddress,showpacs]{revtex4-1}
\usepackage{graphicx}

\begin{document}


\title{Many Topological Insulators Fail the Surface Conduction Test}


\author{Sourabh Barua}
\email[]{sbarua@iitk.ac.in}
\author{K. P. Rajeev}
\affiliation{Department of Physics, Indian Institute of Technology Kanpur, Kanpur 208016}



\begin{abstract}
In this report, we scrutinize the thickness dependent resistivity data from the recent literature on electrical transport
measurements in topological insulators. A linear increase in resistivity with increase in thickness is expected in the
case of these materials since they have an insulating bulk and a conducting surface. However, such a trend is not seen in
the resistivity versus thickness data for all the cases examined, except for some samples, where it holds for a narrow
range of thickness.
\end{abstract}

\pacs{73.25.+i, 73.20.-r, 03.65.Vf }

\maketitle

\section{Introduction}
Topological insulators are the latest phase to have been discovered in condensed matter physics and due to their unique
properties and potential applications they are one of the current hot topics in the field. They are members of a
topological class characterized by the $Z_{2}$ topological invariant and their band structure has extended states
associated with their surfaces within the bulk band gap \cite{Hasan,Qi,Culcer}. The in-gap surface states are
topologically protected by time reversal symmetry and this makes the surface electrically conducting even though the bulk
is insulating. These states at the surface are special in the sense that they have a linear dispersion relation near the
$\Gamma$ point, i.e. $ E \propto |\vec k| $; also the spin of each electron is locked to its wave vector. These
theoretical considerations should lead to the following experimentally observable properties:

\begin{enumerate}
\item Linear energy vs momentum relation of the surface states.
\item A conducting surface and an insulating bulk.
\item Topological protection of current carrying surface electrons from scattering in the absence of magnetic impurities or magnetic fields.
\end{enumerate}

The first property pertains to the band structure while the remaining two are related to electrical transport.
\section{Thickness Dependence of Resistivity: Bulk and Surface Conductors}

\begin{figure}
\includegraphics[width = 0.9\textwidth]{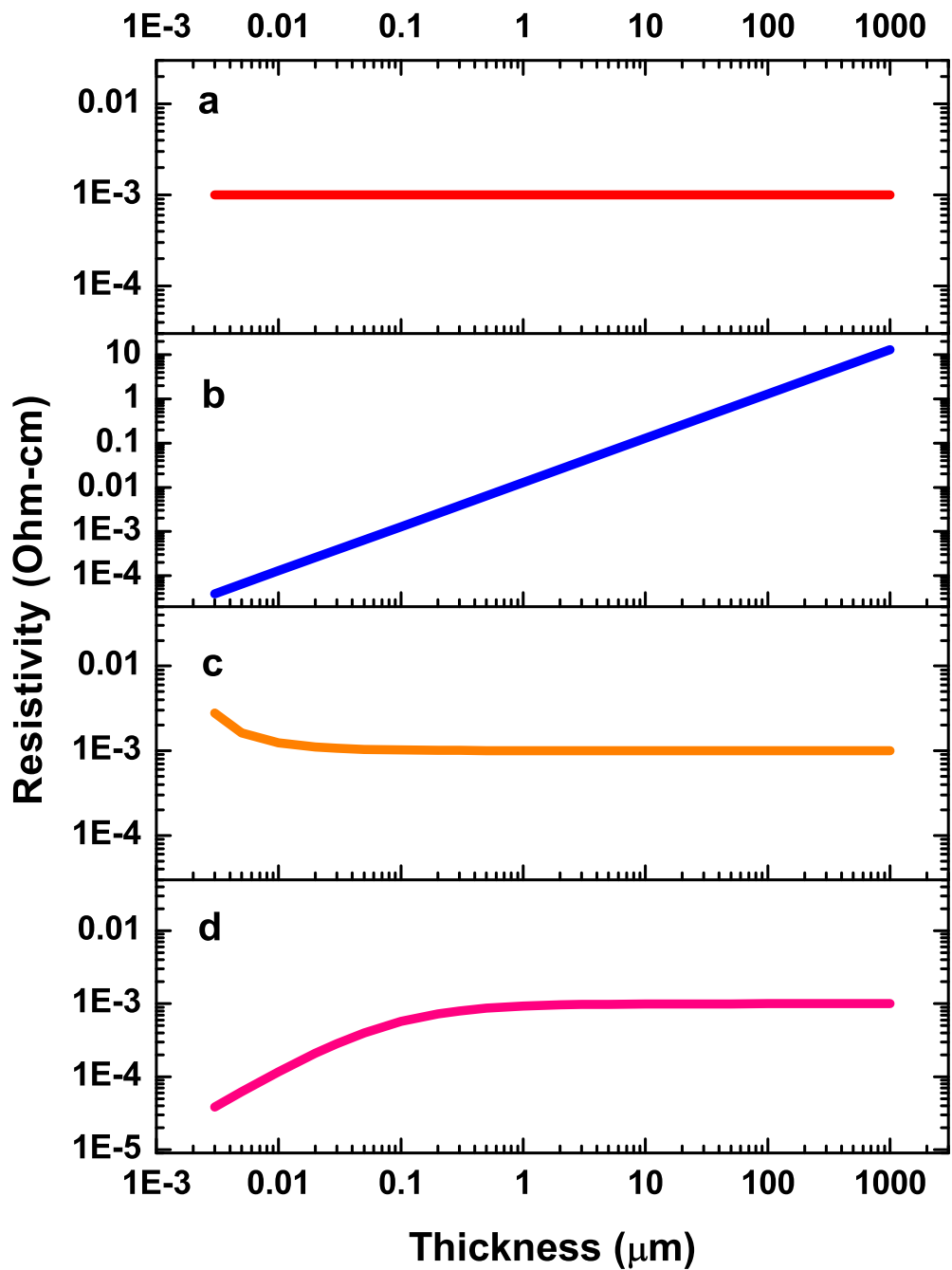}
\caption{ Variation of resistivity with thickness for different degrees of surface and bulk conduction: {\bf(a)} bulk
conduction, {\bf(b)} pure surface conduction, {\bf(c)} surface less conducting than bulk, {\bf(d)} surface more conducting than bulk. For details refer to the appendix. \label{Fig1}}
\end{figure}

Evidence for in-gap surface states, with their linear energy dispersion relation have been provided by angle resolved
photoemission spectroscopy (ARPES) in topological insulator materials such as Bi$_{x}$Sb$_{1-x}$, Bi$_{2}$Te$_{3}$,
Bi$_{2}$Se$_{3}$, Bi$_{2}$Se$_{2}$Te and Bi$_{2-x}$Sb$_{x}$Te$_{3-y}$Se$_{y}$ \cite{Hsieh,Chen,Xia,Bao,Arakane}. It is
expected that the properties, viz, surface conduction and its topological protection will be evident in the electrical
transport measurements. One way in which surface conduction would manifest itself in transport measurements is in
thickness dependent resistivity measurements because a surface conductor differs from a bulk conductor in the manner in
which its resistivity varies as a function of thickness. Resistivity for three dimensional systems is  given by $ \rho =
\frac{Rwt}{l}$, where $R$ is the resistance, $w$ is the width, $l$ is the length and $t$ is the thickness of the
conductor. This definition of resistivity does not apply to a surface conductor as it is a two dimensional system and has
no thickness. However we shall show that one can make use of this definition of bulk resistivity in analyzing and making sense of electrical conduction in a surface conductor. In what follows, we examine the different types of variation of resistivity with thickness expected when the bulk and surface of a system conduct to different extents. In Fig. 1 we show the results of a calculation of the thickness dependence of resistivity in systems with a range of surface and bulk conduction (See appendix for details of calculation). A system with only bulk conduction has a thickness independent resistivity as shown in Fig. 1a. If a system has only surface conduction then the resistivity increases linearly with increasing thickness, as shown in Fig. 1b. In systems where both surface and bulk conduct, the thickness dependence can be of different types. If the surface is more insulating than the bulk then the resistivity decreases with increasing thickness at low thicknesses, as shown in Fig. 1c. If the surface is more conducting than the bulk, then the resistivity increases linearly at low thicknesses and then flattens off as shown in Fig. 1d.

From the above discussion it is clear that the thickness dependent transport measurements could be an important tool to
investigate the extent of surface conduction in topological insulators. In most samples of topological insulators, the
Fermi level lies in a band formed due to defects and vacancies, making most of the as grown crystals
metallic\cite{Culcer,Taskin,Cao} and as a result most of the reports on electrical transport measurements are on such
metallic systems \cite{Cao,Taskin3,Qu,Butch,Kim,Analytis,Bansal,Eto,Analytis2}. Now, a true topological insulator cannot
have a conducting bulk and hence efforts have been made to obtain insulating crystals by various methods such as doping
Bi$_{2}$Se$_{3}$ with calcium\cite{Hor,Checkelsky}, lowering carrier density and partial substitution of bismuth with
antimony in Bi$_{2}$Se$_{3}$\cite{Analytis}, growing Bi$_{2}$Te$_{3}$ single crystals with compositional
gradient\cite{Qu}, and annealing Bi$_{2}$Te$_{3}$ single crystals in tellurium vapour\cite{Hor1}. Also, as grown crystals
of Bi$_{2}$Te$_{2}$Se (BTS) were found to be insulating\cite{Ren,Xiong} and partial replacement of bismuth with antimony
in Bi$_{2}$Te$_{2}$Se led to the formation of insulating crystals of Bi$_{1.5}$Sb$_{0.5}$Te$_{1.7}$Se$_{1.3}$
\cite{Taskin2}. Although thickness dependent resistivity measurements could be the easiest method to detect signatures of
surface states in the conduction, most of the transport measurements on topological insulators rely on the angular
dependence of the period of the Shubnikov-de Haas oscillations
(SdH)\cite{Analytis,Eto,Taskin,Taskin3,Cao,Qu,Taskin4,Analytis2,Ren,Taskin2,Tang,Xiong} or extraction of carrier density
by fitting a two band model to the Hall data\cite{Ren,Bansal,Taskin3}. In fact measuring the thickness dependence of
resistivity becomes even more important in the light of the fact that angular dependence of SdH oscillations similar to
what is expected for surface states of a topological insulator has also been seen in a system which behaves as a bulk
combination of many parallel two-dimensional electron systems \cite{Cao}. In spite of its importance, thickness dependent
transport measurements are rarely reported in topological insulators and hence we felt that it would be in order to
review the existing data from the few such studies in the literature.

\section{Thickness Dependence of Resistivity of Topological Insulators}

\begin{figure}
\includegraphics[width = 1\textwidth]{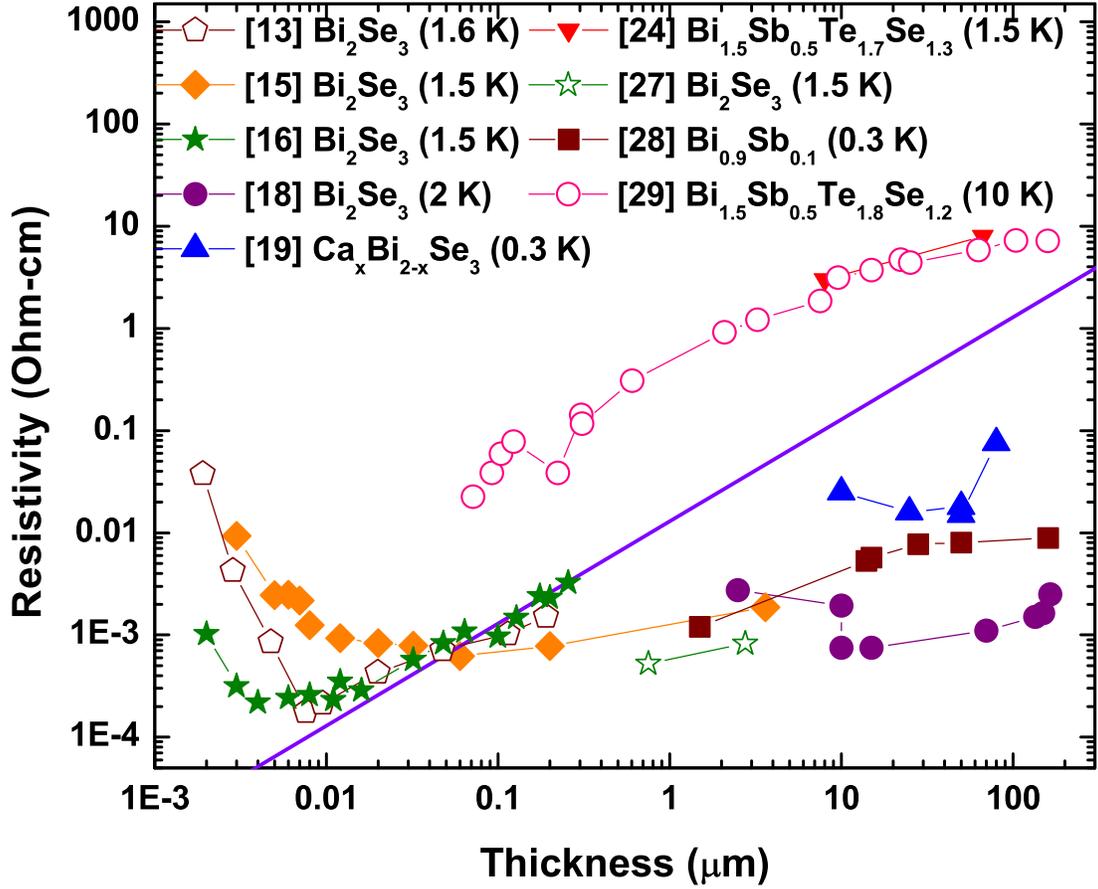}
\caption{\label{Fig2}Resistivity versus thickness for topological insulators from the literature. The legend indicates
the reference number (in square brackets), the name of the sample and the temperature (in parantheses) at which these
measurements were done. The solid line indicates how the resistivity of a true topological insulator will vary with
thickness}

\end{figure}

In Fig. 2 we plot the low temperature resistivity versus thickness data on topological insulators taken from Refs. 13,15,16,18,19,24,27-29. The resistivity at low temperature was chosen so as to rule out thermal effects on conduction. The references from which the different sets of data were taken alongwith the sample names and the temperature at which these measurements were done is indicated in the legend of Fig. 2. The temperature indicated for Ref. 13 is an educated guess as it was not explicitly mentioned in the article. The resistivity and thickness values were directly read off from the plots in Refs. 18,24,29 while it was read off from the tables in Refs. 19,28. In the case of Refs. 13,15,27 the sheet resistance values were read off from the graph and multiplied by the corresponding thickness values to obtain the resistivity data. In the case of Ref. 16 the sheet conductance values were read off from the graph and then its reciprocal taken and multiplied with thickness to obtain the resistivity.

A perfect topological insulator is supposed to have only surface conduction and, as mentioned earlier, the resistivity in
case of pure surface conduction would increase linearly with thickness as shown in Fig. 1b. To serve as a
guide to the eye we have again plotted this linear variation of resistivity with thickness in case of pure surface
conduction as a solid line in Fig. 2. The slope of this line is necessarily unity but its position can shift
as its intercept will depend on the value of surface conductivity chosen. Comparing it with the different sets of
data plotted in Fig. 2, it is obvious that none of the data show a similar linear increase with thickness over
their entire range. This is not contrary to expectation since none of the single crystals of these
materials have a highly insulating bulk. Refs. 13,16 show a near linear increase in resistivity with
increasing thickness at high thicknesses which is very close to that for a pure surface conductor. However, both of them
show a sudden fall in resistivity at low thicknesses which we shall discuss a little later. Ref. 15 also shows a similar low thickness fall, however instead of showing a linear rise at high thicknesses, it remains more or less flat. However a close resemblance between data from Ref. 15 and Fig. 1c is clearly seen implying that this is an example of a case in which both surface and bulk conduct with the surface less conducting. In case of Ref. 28 the resistivity initially increases with thickness but ultimately flattens off. Data from Ref. 29 which is on the sample Bi$_{1.5}$Sb$_{0.5}$Te$_{1.8}$Se$_{1.2}$, whose resistivity is among the highest for topological insulator materials, shows an increase in resistivity with thickness over the largest range of thickness although it is not  strictly linear and also tends to flatten off at high thicknesses. The data from Ref. 28,29 are similar to the case where both the surface and bulk conduct but with the surface more conducting than the bulk as shown in Fig. 1d. A further similarity between our model and the actual data is that the resistivity saturates in case of Fig. 1d at the bulk resistivity value of 1 m$\Omega$-cm which was assumed in our calculation (see Table A.1 in appendix) and in the case of Refs. 28,29 the resistivity saturates at values of approximately 10 m$\Omega$-cm and 10 $\Omega$-cm respectively which are close to the bulk resistivity values reported in those cases. Refs. 18,19 have fewer points and there is no definite trend although there is a hint of a fall in resistivity at low thicknesses as well as a rise at high thicknesses but overall they remain flat and can be considered as bulk conductors.

Coming back to the case of low thickness fall in resistivity, it has been reported that a gap opens up in the band
structure of the surface states 13,30 when the thickness of the topological insulator sample decreases below
a critical thickness and this could be the reason for the behavior seen in Refs. 13,16. The fall in
resistivity occurs only below a thickness of $10$ $nm$ in Ref. 16 which is close to the critical
thickness of $6$ nm \cite{He,Taskin3}. It is unlikely that the low thickness fall in resistivity in Refs.
18,19 is due to this same effect since this happens at much larger thicknesses. The fact that
our calculation also captures this low thickness fall in resistivity when the surface is less conducting than the bulk,
as evident from the graph in Fig. 1c, emphasizes the robustness of thickness dependent resistivity
measurements as a test for surface conduction in topological insulators. Based on the variation in resistivity with
thickness it can be said that only Refs. 13,16,28,29 come close to showing the surface
conduction which is a primary property of topological insulators. However if the data of Ref. 16 is
taken in conjunction with that of Ref. 27, which was an earlier report on the same sample by the same
authors, then the resistivity seems to be flattening off at high thicknesses. Further, since Ref. 28 has
few data-points in the linear portion, we can say that only Refs. 13,29 show unambiguous signs of
surface conduction. Bi$_{1.5}$Sb$_{0.5}$Te$_{1.8}$Se$_{1.2}$ \cite{Bin_Xia} seems to be the most promising candidate for
topological insulator from this thickness dependent study as it shows a nearly linear dependence of resistivity on
thickness for the widest range of thicknesses and also it has the highest resistivity among the materials reviewed in
this report which is ideal since it implies that the resistivity will saturate at higher thicknesses. Thus, even though
surface conduction in topological insulators has been claimed to be detected in transport measurements, not all
topological insulator samples pass the surface conduction test of thickness dependence.

\section{Conclusion}

In conclusion, the thickness dependent study of data on the electrical resistivity of topological insulators in the literature reveals that while some materials show surface conduction, many are simply bulk conductors. At low thicknesses where a gap opens up in the surface states a fall in resistivity with increasing thickness is also seen. Thickness dependence of resistivity can be an important tool to detect surface conduction and find materials which are the best candidates for topological insulators.

\appendix*
\setcounter{table}{0} \renewcommand{\thetable}{A.\arabic{table}} 

\section{Appendix: Variation of resistivity with thickness}
\begin{figure}[h!]
\includegraphics[width = 1\textwidth]{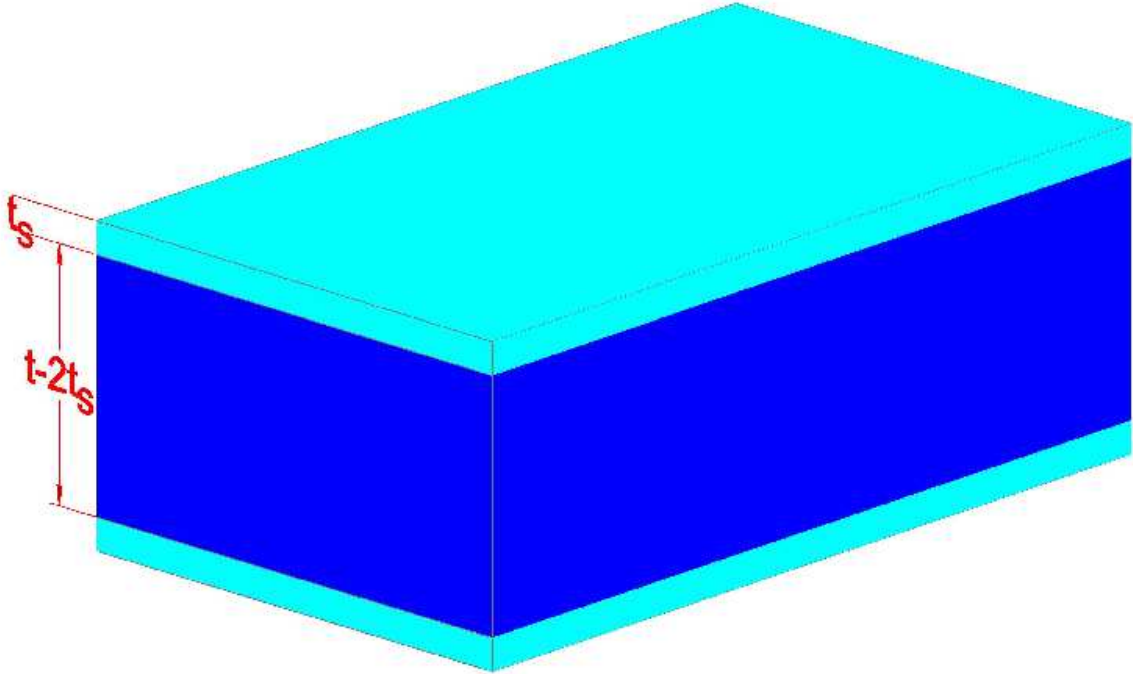}
\caption{\label{Fig_Appendix}The model used to calculate thickness dependence of electrical resistivity in systems with
both bulk and surface conduction. $t$ is the total thickness of the conductor and $t_{s}$ is the thickness of each of the
two surfaces; top and bottom.}

\end{figure}

In Fig. 3 we show the model used to obtain the thickness dependence of resistivity when the surface
and bulk of a system conduct to different extents. Resistivity in case of a pure bulk conductor is independent of
thickness and hence in Fig. 1A the resistivity is shown to be flat with thickness at a constant value of 1
m$\Omega$-cm which we have assumed for the bulk resistivity value. In case of only surface conduction, if we assume a
bulk with infinite resistivity, then the resistivity is simply given by $ \rho = \frac{R_s\times t}{2}$, where t is the
thickness of the sample and $R_s$ is the resistance of one surface. The factor of 2 comes because we consider two
surfaces in parallel. Here, we assumed $R_s$ to be 258.13 $\Omega $. In Ref. \cite{Qu} the surface conductance of a
topological insulator was estimated to be of the order of 100 $e^{2}/h$ which is equivalent to a resistance of 258.13
$\Omega$. For obtaining the variation of resistivity with thickness in case of conductors with both bulk and surface
conduction, we have modeled the conductor as a parallel combination of one bulk conductor sandwiched between two surface
layers. The surface layers have been assumed to have a thickness of 1 nm and a surface resistance $R_s$. The bulk
resistivity $\rho_b$ and surface resistance $R_s$ values were chosen in such a manner that we could simulate different
degree of bulk and surface conduction. The thickness of the surface layers has been kept constant in all the different
cases. We obtain the following expression for the resistivity $\rho$ of the whole conductor,

\begin{equation}
\label{app_equation} \rho = \frac{\rho_b \rho_s t}{2\rho_b t_s + \rho_s(t - 2 t_s)}
\end{equation}

\begin{table}[h!]
\begin{tabular}{|l|l|l|}
\hline Type of conduction & $R_s$ ($\Omega$) & $\rho_b$ ($\Omega$-cm) \\ \hline Surface less conducting than bulk &
258130 & 0.001 \\ \hline Surface more conducting than bulk & 258.13 & 0.001 \\ \hline
\end{tabular}
\caption{Table of values of $R_s$ and $\rho_b$ used in the calculation.} \label{appTable}
\end{table}

where $ t_s$ is surface thickness, $t$ is the thickness of the sample and $\rho_s$ is given by $R_s \times t_s$.
Different values of $R_s$ and $\rho_b$ were used to simulate the different degrees of bulk and surface conduction and are
given in Table 1.

\begin{acknowledgments}
 SB acknowledges CSIR, India for financial support.
\end{acknowledgments}


\end{document}